\documentclass[aps,pra,twocolumn,floatfix,nofootinbib]{revtex4-2}
\usepackage{amsmath}
\usepackage{graphicx}
\usepackage{amssymb}
\usepackage{amsmath}

\begin{document}
\title{Efficiently computing the Uhlmann fidelity for density matrices}
\author{Andrew J. Baldwin}\email{andrew.baldwin@chem.ox.ac.uk}
\affiliation{Chemistry Research Laboratory, University of Oxford, Mansfield Road, Oxford OX1 3TA, UK}
\affiliation{Kavli Institute for Nanoscience Discovery, University of Oxford, Oxford OX1 3QU, UK}
\author{Jonathan A. Jones}\email{jonathan.jones@physics.ox.ac.uk}
\affiliation{Clarendon Laboratory, University of Oxford, Parks Road, Oxford OX1 3PU, UK}
\date{\today}

\begin{abstract}
We consider the problem of efficiently computing the Uhlmann fidelity in the case when explicit density matrix descriptions are available. We derive an alternative formula which is simpler to evaluate numerically, saving a factor of 10 in time for large matrices.
\end{abstract}

\maketitle

\section{Introduction}
The Uhlmann--Jozsa fidelity \cite{Uhlmann1976, Jozsa1994} between two general density operators, $\rho$ and $\sigma$ is defined as
\begin{equation}
F_U=\left[\textrm{Tr}\left(|\sqrt{\rho}\sqrt{\sigma}|\right)\right]^2,\qquad |A|=\sqrt{AA^\dag}.\label{eq:FUJ}
\end{equation}
Note that all proper density operators are Hermitian and positive semi-definite with trace 1, and so $\sqrt\rho$ and $\sqrt\sigma$ exist. They are also Hermitian and positive semi-definite, but do not have trace 1 except in the special case of a pure state. This leads to the more usual form
\begin{equation}
F_U=\left[\textrm{Tr}\left(\sqrt{\sqrt{\rho}\,\sigma\sqrt{\rho}}\right)\right]^2.\label{eq:Fclassic}
\end{equation}
This all works because
\begin{equation}
M=\sqrt{\rho}\,\sigma\sqrt{\rho}
\end{equation}
is Hermitian and positive semi-definite by construction, and so $\sqrt{M}$ exists and is also Hermitian and positive semi-definite, with eigenvalues equal to the square root of the eigenvalues of $M$.

This definition has the advantage of both obeying reasonable axioms for a fidelity and also having an underlying physical interpretation \cite{Jozsa1994} in terms of the maximal overlap of all purifications of $\rho$ and $\sigma$, but has the disadvantage of being relatively expensive to compute. As a consequence many alternative fidelities have been proposed \cite{Mendonca2008, Liang2019} for mixed states. Here we adopt a simpler approach, recasting the fidelity formula in a way which is easier to calculate numerically.

\section{Proof}
Our key observation is that $F_U$ can be calculated from the eigenvalues of $M$,
\begin{equation}
F_U=\left[{\textstyle\sum_j}\sqrt{\lambda_j}\right]^2,\label{eq:Feigenvalues}
\end{equation}
and so can instead be calculated from any other matrix which is similar to $M$, that is any matrix of the form
\begin{equation}
N=PMP^{-1},\label{eq:sim}
\end{equation}
where $P$ is some invertible transformation. Note that we can write
\begin{equation}
\sqrt{N}=P\sqrt{M}P^{-1}
\end{equation}
with the same transformation, and so $\sqrt{N}$ will be similar to $\sqrt{M}$, and so have the same eigenvalues \cite{RHB2006}.

In the case when $\rho$ is positive definite we can choose $P=\sqrt\rho$, which will have an inverse. Then
\begin{equation}
N=PMP^{-1}=(\sqrt\rho)(\sqrt\rho\,\sigma\sqrt\rho)(\sqrt\rho)^{-1}=\rho\sigma.
\end{equation}
will be suitable, leading to the alternative form
\begin{equation}
F_U=\left[\textrm{Tr}\left(\sqrt{\rho\sigma}\right)\right]^2\label{eq:Fnew}
\end{equation}
as a convenient way of computing the Uhlmann--Jozsa fidelity for density matrices.

If $\rho$ is positive semi-definite but not positive definite then this approach cannot be used as $\sqrt\rho$ does not have an inverse. However this can be sidestepped. Working in the eigenbasis of $\rho$ and separating the positive definite block (dimension $p$) from the null block (dimension $q$) gives
\begin{equation}
\begin{split}
M=\sqrt{\rho}\,\sigma\sqrt{\rho}
&=
\begin{pmatrix}\sqrt{\rho_p}&0_c\\0_c^\dag&0_q\end{pmatrix}
\begin{pmatrix}\sigma_p &\sigma_c\\\sigma_c^\dag&\sigma_q\end{pmatrix}
\begin{pmatrix}\sqrt{\rho_p}&0_c\\0_c^\dag&0_q\end{pmatrix}\\
&=
\begin{pmatrix}\sqrt{\rho_p}\,\sigma_p\sqrt{\rho_p}&0_c\\0_c^\dag&0_q\end{pmatrix}
=
\begin{pmatrix}M_p&0_c\\0_c^\dag&0_q\end{pmatrix},
\end{split}
\end{equation}
where $0_q$ is a $q$-by-$q$ square block of zeros and $0_c$ is a rectangular block of zero-valued cross terms. Thus $M$ has the same block structure as $\rho$. Similarly
\begin{equation}
\begin{split}
N=\rho\sigma
&=
\begin{pmatrix}\rho_p&0_c\\0_c&0_q\end{pmatrix}
\begin{pmatrix}\sigma_p &\sigma_c\\\sigma_c^\dag&\sigma_q\end{pmatrix}\\
&=
\begin{pmatrix}\rho_p\sigma_p&\rho_p\sigma_c\\0_c&0_q\end{pmatrix}
=
\begin{pmatrix}N_p&N_c\\0_c&0_q\end{pmatrix},
\end{split}
\end{equation}
which does not have the same structure but is still block upper triangular in this basis with the same null block.

The determinant of a block triangular matrix equals the product of the determinants of the individual diagonal blocks \cite{Silvester2000}, and so the characteristic polynomial, and thus the eigenvalues, are unaffected by the values in $N_c$. Hence $N$ has the same eigenvalues as $M$ if $N_p$ is similar to $M_p$. For the $M_p$ block we can now choose $P_p=\sqrt{\rho_p}$, which has an inverse as $\rho_p$ is positive definite by construction. Thus equation~\ref{eq:Fnew} can still be used when $\rho$ is only positive semi-definite.

An equivalent approach uses a Drazin pseudo-inverse \cite{Drazin1958} in place of the inverse in equation~\ref{eq:sim}. In the eigenbasis of $P$ this has eigenvalues equal to the reciprocals of the original eigenvalues, except that zero eigenvalues are instead mapped to zeros. Thus the product of $P$ and its Drazin pseudo-inverse is not the identity, but instead a projector onto the non-null subspace of $P$. As such projections do not change the eigenvalues of a matrix product involving $P$ this permits results which hold for positive definite matrices to be extended to positive semi definite matrices, as was done above.

One could instead use $F_U=\left[\textrm{Tr}\left(\sqrt{\sigma\rho}\right)\right]^2$. The proof in this case is identical except that $N$ now has block lower triangular form. More generally the eigenvalues of any sandwich $\rho^a\sigma^b\rho^c$ are unaffected by permuting the outer layers, or  any portion thereof, and in particular are the same for the open-face sandwich $\rho^{a+c}\sigma^b$. Thus when evaluating $F_U$ any matrix of the form $\rho^x\sigma\rho^{1-x}$ can be used, with the case $x=1/2$ corresponding to the conventional form, equation~\ref{eq:Fclassic}, and the cases $x=1$ and $x=0$ corresponding to the two cases considered above. However choosing any value of $x$ outside these three special cases has no obvious advantage. A related approach based on the properties of eigenvalues has been explored by Wang and Gong \cite{Wang1993}.

\section{Efficiency}
We have now seen three different forms which permit $F_U$ to be computed for any proper density matrices: two traditional forms, equations \ref{eq:FUJ} and \ref{eq:Fclassic}, and our new form, equation~\ref{eq:Fnew}. All three forms will give identical results, but they will take different effort to compute. We take equation \ref{eq:Fclassic} as the base case, which requires two computationally expensive matrix square-roots, two matrix products, and a trace operation. This is clearly faster than equation \ref{eq:FUJ}, which requires three matrix square-roots. The new form, equation~\ref{eq:Fnew}, is even better, using only a single matrix square-root, but the most efficient route replaces this final expensive operation with a significantly faster eigenvalue calculation and the use of equation \ref{eq:Feigenvalues}.

As the time required for the key steps in both the base case and our new efficient approach scales with the dimension $n$ of the density matrices as $O(n^3)$, the new approach should be faster for large matrices than the conventional approach by a constant factor, which we estimate from numerical studies to be around 10. 

Similar ideas have been explored in the context of R\'enyi relative entropies \cite{Audenaert2015}, and similar efficiencies should be achievable by replacing sandwiches with their open-face equivalents.

\section{Corollaries}
The form above can be used to rederive several well known properties \cite{Jozsa1994} of the Uhlmann--Jozsa fidelity. It is obvious from this form that $F_U$ is symmetric to interchange of $\rho$ and $\sigma$, that $F_U$ is invariant under bilateral unitary transformations, since
\begin{equation}
U\rho U^\dag\, U\sigma U^\dag=U\rho\sigma U^\dag
\end{equation}
is clearly similar to $\rho\sigma$, and that $F_U=1$ when $\rho=\sigma$ since $\sqrt{\rho\rho}=\rho$ and $\textrm{Tr}(\rho)=1$. Using the identities
\begin{equation}
\sqrt{(\rho_1\otimes\rho_2)(\sigma_1\otimes\sigma_2)}=\sqrt{\rho_1\sigma_1}\,\otimes\sqrt{\rho_2\sigma_2},
\end{equation}
and
\begin{equation}
[\textrm{Tr}(A\otimes B)]^2=[\textrm{Tr}(A)]^2\times[\textrm{Tr}(B)]^2,
\end{equation}
it also follows that $F_U$ is multiplicative, that is
\begin{equation}
F_U(\rho_1\otimes\rho_2,\sigma_1\otimes\sigma_2)=F_U(\rho_1,\sigma_1)F_U(\rho_2,\sigma_2).
\end{equation}

A final result comes from expanding the eigenvalue expression, equation~\ref{eq:Feigenvalues},
\begin{equation}
\begin{split}
F_U&=\sum_{j,k}\sqrt{\lambda_j}\sqrt{\lambda_k}\\
&=\sum_{j}\lambda_j+2\sum_{j<k}\sqrt{\lambda_j}\sqrt{\lambda_k}
\end{split}
\end{equation}
where the first term is recognisable as $\textrm{Tr}(M)=\textrm{Tr}(N)$, and the (non-negative) eigenvalues in the second term can be calculated from $N$ rather than $M$. Thus
\begin{equation}
F_U=\textrm{Tr}(\rho\sigma)+2\sum_{j<k}\sqrt{\lambda_j\lambda_k},
\end{equation}
as previously noted by Miszczak \textit{et al.}~\cite{Miszczak2009}. The first term is recognisable as the fidelity between a pure state $\rho=|\psi\rangle\langle\psi|$ and a general state $\sigma$, as
\begin{equation}
F=\langle\psi|\sigma|\psi\rangle=\textrm{Tr}(|\psi\rangle\langle\psi|\sigma)=\textrm{Tr}(\rho\sigma).
\end{equation}
The second term, which is non-negative, is calculated from the eigenvalues of $N=\rho\sigma$. (An equivalent result for $M$ was previously noted for a single qubit \cite{Jozsa1994, Hubner1992}.) If $\rho$ is pure then it has rank 1, that is it has precisely one non-zero eigenvalue, and the same is true for $\sigma$. Finally the rank of a matrix product is no larger than the smaller of the ranks of the two components \cite{Meyer2000}, and thus $\rho\sigma$ is at most of rank 1 if either $\rho$ or $\sigma$ is pure. In this case all the terms in the sum vanish, and we are left with
\begin{equation}
F_U=\textrm{Tr}(\rho\sigma)
\end{equation}
as expected.

\begin{acknowledgments}
Andrew Baldwin is supported by ERC grant (101002859). For the purpose of Open Access, the author has applied a CC BY public copyright license to any Author Accepted Manuscript version arising from this submission. We thank Artur Ekert, Karol \.{Z}yczkowski and Mark Wilde for helpful conversations.
\end{acknowledgments}

\bibliography{Uhlmann}
\end{document}